\documentclass{article} % For LaTeX2e
\usepackage{iclr2019_conference,times}
\usepackage[utf8]{inputenc}

\usepackage{graphicx}% Include figure files
\usepackage{dcolumn}% Align table columns on decimal point
\usepackage{bm}% bold math
\usepackage[mathlines]{lineno}% Enable numbering of text and display math
\usepackage{xcolor}
\usepackage{hyperref}
\usepackage{url}
\usepackage{breakurl}
\usepackage{graphicx}
\usepackage{caption}
\usepackage{amsmath}
\usepackage{cleveref}
\usepackage{textgreek}
\usepackage{xspace}
\usepackage[binary-units]{siunitx}
\usepackage{booktabs}
\usepackage{listings}
\usepackage{xcolor}
\usepackage{verbatim}
\usepackage{subfigure}
\usepackage{amsfonts}

\usepackage{listings}
\title{Using reinforcement learning to minimize taxi idle times}

\iclrfinalcopy

\author{Kevin O'Keeffe \\
Senseable City Lab, Massachusetts Institute of Technology \\
Cambridge, MA 02139 \\
\texttt{kokeeffe@mit.edu} 
\And
Sam Anklesaria \\
Senseable City Lab, Massachusetts Institute of Technology \\
Cambridge, MA 02139 \\
\And
Paolo Santi \\
Senseable City Lab, Massachusetts Institute of Technology \\
Cambridge, MA 02139 \\
\And 
Carlo Ratti \\
Senseable City Lab, Massachusetts Institute of Technology \\
Cambridge, MA 02139 \\
}

\date{\today}

\begin{document}
\maketitle

\maketitle

\begin{abstract}
Taxis spend a significant amount of time idle, searching for passengers. The routes vacant taxis should follow in order to minimize their idle times are hard to calculate; they depend on complex quantities like passenger demand, traffic conditions, and inter-taxi competition. Here we explore if reinforcement learning (RL) can be used for this purpose. Using real-world data to characterize passenger demand, we show RL-taxis indeed learn to how to reduce their idle time in many environments. In particular, a single RL-taxi operating in a population of regular taxis learns to out-perform its rivals by a significant margin.
\end{abstract}

%Taxis spend a significant amount of time idle, searching for passengers. The optimal routes vacant taxis should follow in order to minimize their idle times are hard to calculate; they depend on complex quantities like passenger demand, traffic conditions, and inter-taxi competition. Here we explore if reinforcement learning (RL) can be used to find `optimal paths' for vacant taxis -- the paths which when followed minimize idle times. Using real-world data to characterize passenger demand, we show RL-taxis indeed learn to how to reduce their idle time in a variety of environments. In particular, a single RL-taxi operating in a population of regular taxis learns to out-perform its rivals by a significant margin.

%%%%%%%%%%%%%%%%%%%%%%%%%%%%%%%%%%%%%%%%%%%%%%%%%%%%%%%%%%%%%%%%%%%%%%%%%%%%%%%%%%%%%%%%%%%%55

\section{Introduction}
Even during peak hours, taxis in Manhattan spend one third of their driving time idle \cite{schaller2017}, looking for passengers. Estimates of idle times for taxis in other cities are as high as 60\% \cite{powell2011towards}. This inefficiency bears much cost: it worsens congestion for commuters, reduces revenue for taxi companies, and increases pollution to the environment. A reduction in taxi idle times would thus confer much benefit.

%Why this is a hard problem. What others have done.
Reducing taxi inefficiency amounts to an optimal path finding problem: when vacant, what path should a taxi follow to minimize its idle time, the total time it spends looking for passengers? Most previous research found step-wise optimal solutions to this problem \cite{zhang2014online,qu2014cost,dong2014recommend,huang2014backward} -- `greedy' paths which minimize the time until the \textit{next} passenger pick-up only. But because the idle time is a cumulative quantity calculated over a given reference period -- and therefore depends on the \textit{full} trajectory a taxi traces out in both serving and searching for multiple passengers -- these greedy paths are not necessarily globally optimal. A recent work \cite{yu2019markov} has improved upon this situation by formulating the problem as a Markov Decision Process (MDP), the natural framework to optimize cumulative, and not just instantaneous, quantities / objectives. Using taxi data from Shanghai, they showed this formalism performed better than common baselines.

%Problem with MDP
%In order for MDPs to be effective, however, an accurate model of the environment must be provided. For taxis moving in cities, this is a challenging task. The relevant features, such as passenger demand, traffic conditions, and inter-taxi competition, are complicated quantities with potentially non-trivial interactions. As such, these effects comprise a complex system -- which, by definition, are difficult to characterize. The benefit of using MDPs to minimize taxi idle times could thus be limited. 

In order for MDPs to be effective, however, an accurate model of the environment must be provided. For taxis moving in cities, this is a challenging task. The relevant features, such as passenger demand, traffic conditions, and inter-taxi competition, are complicated , hard-to model quantities. The ability of MDPs to minimize the idle times of real-world taxis could thus be limited. 

%RL pitch
Reinforcement learning (RL) could overcome this limitation. It is a type of MDP where, ideal for taxis moving in cities (and other complex systems), a model of the environment is not required. Instead, the optimal behavior for the (unspecified) environment is learned through trial and error. When combined with modern techniques from deep learning, so called deep-RL algorithms have been shown to be dramatically powerful. They have been used to train autonomous helicopters \cite{ng2006autonomous,abbeel2007application}, to achieve super human performance in the board game Go \cite{silver2017mastering}, and to train a robotic hand to solve a Rubix cube \cite{andrychowicz2018learning}.

%What we do
It's natural to wonder if these successes could be translated to taxi research. Several researchers have started to explore this possibility by applying deep-RL in various contexts, such as optimizing taxi carpooling \cite{jindal2018optimizing}, dispatching \cite{oda2018movi}, and route scheduling \cite{shi2018deep}. In this work, we continue this endeavour by studying if deep-RL could be useful in minimizing taxi idle times. A first step in this direction was taken in \cite{han2016routing}, where a single RL-taxi on an otherwise empty street network from Singapore was trained using tabular $Q$-learning (a basic form of RL which doesn't rely on deep learning). Here we extend this work by using more powerful RL algorithms in new environments (i) a single RL-taxis moving on street networks from several real world cities, and (ii) a single RL-taxi moving in a population of regular (non-RL) taxis. We find RL-taxis can indeed learn to appreciably lower their idle times, as compared to other baseline taxi models. Our results are potentially relevant to taxi companies, intelligent transport systems, and other settings in which vehicular efficiency is important. %We hope our work inspire further research in applying RL to taxis and other mobility systems.

%%%%%%%%%%%%%%%%%%%%%%%%%%%%%%%%%%%%%%%%%%%%%%%%%%%%%%%%%%%%%%%%%%%%%%%%%%%%%%%%%%%%%%%%%%%%$$

\section{Results}

\subsection{Background materials}

\textbf{Mathematical preliminaries}. We begin with a definition of an MDP, which models an agent interacting with an environment. The agent's interactions leads to rewards, which the agents tries to maximize. Mathematically, an MDP is defined by a 4-tuple $(\mathcal{S}, \mathcal{A}, P_a, R_a)$, where $\mathcal{S}$ is a set of states which characterize the environments, $\mathcal{A}$ is a set of actions available to the agent, $P_a(s,s')$ is a transition function, which describes the probability of the environment going from state $s$ to $s'$ when the agent takes action $a$, and $R_a(s,s')$ is a function which generates a reward $r$ during this transition from $s$ to $s'$. The goal of an MDP problem is for an agent to learn a policy $\pi(a,s)$ which describes the probability of selecting action $a$ when in state $s$. More specifically, the goal is to learn an optimal policy $\pi^*$ which maximizes the cumulative reward (sometimes called the \textit{return}) $G = \sum_i^T r_i$, where $T$ is the length of the `episode' (the total number of steps or time units considered).

Having defined an MDP we can now define RL: As mentioned in the Introduction, it is simply `model-free' MDP. In other words, an MDP where the system transition function $P_a$ is unknown. Put broadly, then, the purpose of RL is to approximate the optimal policy $\pi^*$ when $P_a$ is either unknown or too complicated to describe analytically. Many algorithms to find $\pi^*$ exists, and finding new algorithms is an active area of research. Generally speaking, the algorithms fall into two classes (a) policy gradient methods and (b) value based methods. In (a) the policy $\pi_{\theta}(s,a)$ is modeled directly by some function approximator $f_{\theta}(s,a)$ (in deep-RL this is a neural network) with parameters $\theta$. This is updated / improved directly via gradient descent methods; that is, by relating $\nabla_{\theta} f$ to $\nabla_{\theta} G$ (intuitively, by moving in the direction in $\theta$-space which maximizes the return $G$). In (b), the policy is defined indirectly. First, a $Q$-function is learned, which is defined as the expected return of taking action $a$ and when in state $s$ : $Q(s,a) = \mathbb{E}(G) = \mathbb{E}(\sum_{i}^{T} r_i)$. The optimal policy is then given by $ \pi(s,a) = argmax_{a'} Q(s,a')$. In deep-RL the $Q$-function is modeled with a neural network. In contrast to policy gradients methods (a), the $Q$ function is updated recursively via the Bellman equation: $Q(s,a) = r + \max_{a'} Q(s,a')$, which can be derived by conditioning on the expectation in the definition of $Q(s,a)$. For a more detailed explanation of these methods, and RL generally, see \cite{sutton2018reinforcement}.

\textbf{Problem formulation}.
We are now ready to formulate the taxi idle time minimization problem as an MDP. We consider taxis moving in a street network $S$ whose edges represent street segments, and whose nodes represent intersections.  We assume each street segment has unit length and taxis travel with constant speed equal to one segment per unit time. The state space $\mathcal{S}$ are the set of positions the taxis can take, or the nodes in the graph. As is common in RL studies, we represent each state by a one-hot vector $s = (0,1,0,0, \dots, N_{nodes})$ where $N_{nodes}$ is the number of nodes in the graph and $s_j = 1$ if the taxi is at node $j$ (although for some environments we consider, different state vectors will be chosen). The action space is $\mathcal{A} = (1,\dots k_{max})$, $k_{max}$ is the maximum node degree in the street network. Selecting action $a = j$ means the taxi moves to the $j$'th neighbour of the current node. Because the number of neighbour a given node is variable, not all action will be `legal' -- selecting $a = 5$ on a node with only two neighbours will be meaningless. In this case, the taxi stays at the current positions and received a negative reward $r = -r_{penalty} = -2$, which will discourage the agent taking this action in the future. In the case where a legal action is taken, the taxi moves to the next node. If a passenger trip is generated there (which happens with a probability specified later), a positive reward $r_{success} = L$, is generated where $L$ is the duration of the trip. If no trip is generated, a reward $r = 0$ is generated. Finally, as discussed the taxi moves according to a policy $\pi(s,a)$, whose optimization (in terms of minimizing the idle time), is the goal of the paper. 

We next consider two specific environments.

%%%%%%%%%%%%%%%%%%%%%%%%%%%%%%%%%%%%%%%%%%%%%%%%%%%%%%%%%%%%%%%%%%%%%%%%%%%%%%%%%%%%%%%%%%%%%%%%%%%%%%%%%%%%%%%%%%%%%%%%%%%%%%%%%%%%%%%%

\subsection{Environment 1: Single taxi}

%%%%%% Figure 1
\begin{figure*}[t!]
  \includegraphics[width= \linewidth]{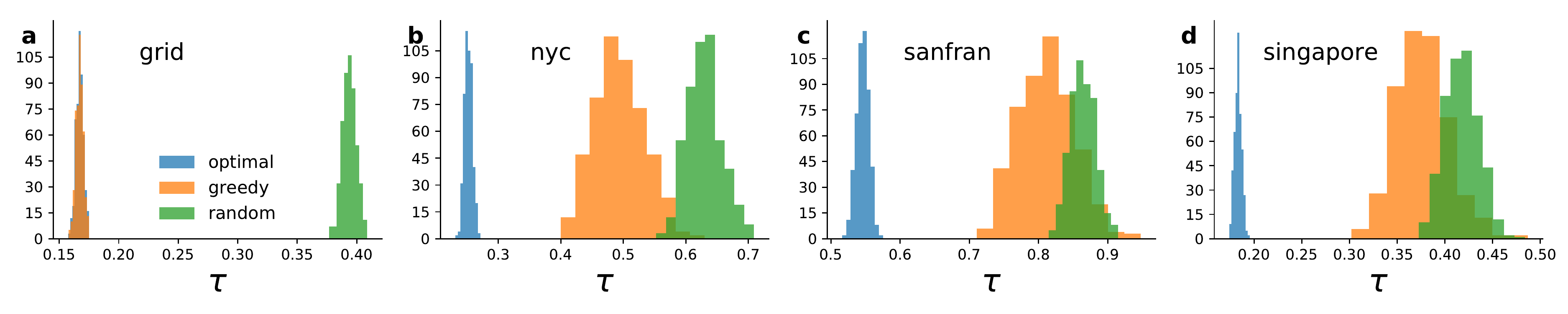}
  \caption{Histogram of idle times $\tau$ for a taxi with policies $\pi^*$, $\pi_{greedy}$, and $\pi_{random}$ on four street networks studied. The optimal policy $\pi^*$ gives significantly lower idles times than the other policies. The ensemble size for each plot was $1000$.}
  \label{idle_times}
\end{figure*}

\textbf{Model environment $\mathcal{M}$}. We begin with the simplest environment $\mathcal{M}$: a single taxi moving on an otherwise empty street network (i.e. no other taxis present). We make the further simplification of there being a constant probability $p_i$ of a passenger being generated at the $i$-th node. This amounts to specifying the transition function $P_a$ -- to providing a model for the environment -- which in turn reduces the problem to a classic MDP. So strictly speaking RL is not needed in this case. Our reason for studying this simple scenario is that the optimal $\pi^*$ can be derived explicitly in this case, and thus can be used as a baseline to test the RL algorithms against, as well as for the more complicated $N$ taxi case. Finally, we also assume street segments have unit length and that passengers persist for precisely one time unit at each node (in Appendix~\ref{appendix:general} we show how these simplifications can be relaxed).

We now derive $\pi^*$. Let the random variable $X_i$ be the time taken for an empty taxi to find the next passenger having started at node $i$. Let the indicator random variable $I_i$ be $1$ if our taxi picks up a passenger at street $i$ within one time-step. If all streets take $1$ time-step to traverse and we continue driving until we find a passenger
\begin{equation}
X_i = I_i + (1 - I_i)(1 + \min_{j \in N_o(i)}X_j)
\label{first}
\end{equation}
where $N(i)$ gives the outgoing neighbors of street $i$. Applying the expectation operator to the above gives
\begin{equation}
x_i = p_i+(1-p_i)\left(1+\min_{j\in N(i)}x_j\right)
\label{relax}
\end{equation}
where $x_i$ is the expected waiting time for a taxi on street $i$. Eq.~\ref{relax} can be interpreted as a relaxation algorithm for $x_i$ -- in other words, can be solved recursively to find $x_i$. Let $x_i^t$ be our approximation of $x_i$ at time $t$. Then we can find a better approximation via $x_i^{t+1} = p_i+(1-p_i)\left(1+\min_{j\in N(i)}x_j^t\right)$. Starting from arbitrary values for $x_i^0$, we continue iterating until we reach a fixed point. Once these $x_i$ are known, the optimal route is found by moving to minimize waiting time: when at node $i$, node $j^* = argmin_{N_i} x_j$ is selected. The optimal policy then corresponds to the optimal route: $\pi^*(j^*|i) = 1 $ which implies $\pi(j|i) = 0$ for all other $j \in N_i$. Note we have abused notation slightly here by using $\pi(j|i)$ to denote the probability of moving to node $j$ when at node $i$ (this is abusive since we earlier defined the policy is term of a state $s$ and action $a$, $\pi(s,a)$, so by $\pi(j|i)$ we mean $\pi(s = i,a')$ where $a'$ denotes to node $j$ is selected when at node $i$).

We tested $\pi^*$ using real-world taxi data from three cities: NYC (confined to Manhattan), San Francisco, and Singapore. We also considered a toy `grid' street network consisting of a square grid of length $n = 5$ (so 25 nodes total), making four cities in total. In Appendix~\ref{appendix:ds} we describe the datasets and how they were used to approximate the $p_i$. 

The street networks themselves were found using the python package `osmnx'. Figure~\ref{idle_times} shows the distribution of idle times of taxis following $\pi^*$. Two baselines are also shown: the idle times for taxis following a random policy $\pi_{random}$, in which a taxi moves randomly on the graph, and a greedy policy $\pi_{greedy}$, in which when at node $i$, the taxi moves to neighbouring node with highest $p_i$. As can be seen, the optimal policy gives substantially lower idles times.

%%%%%% Figure 2
\begin{figure*}[t!]
  \includegraphics[width= \linewidth]{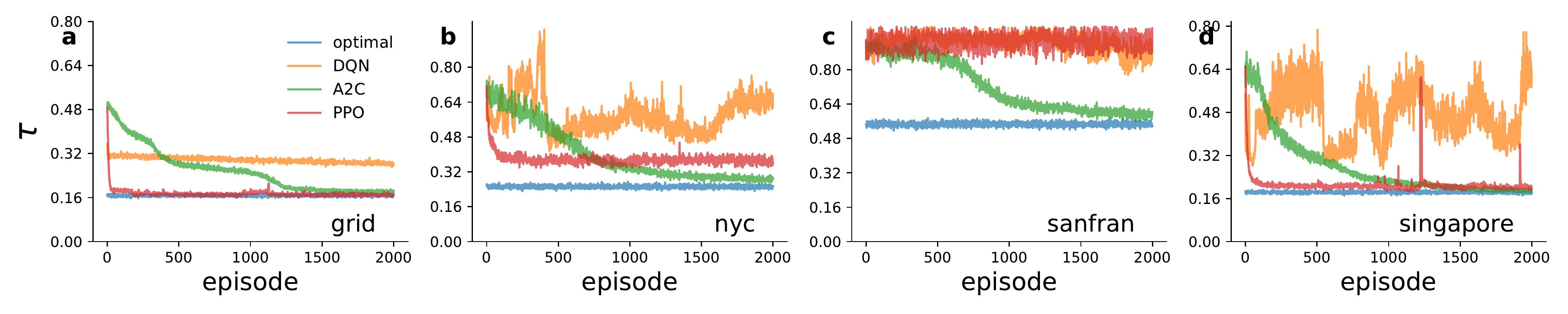}
  \caption{Training curves for RL-taxis for the three RL algorithms used on the four street networks studied.The hyperparameters for these algorithms are report in Appendix~\ref{appendix:ho}.}
  \label{taus-RL}
\end{figure*}

Next we explored how well $\pi^*$ can be approximated by an RL-taxi. We use three popular deep-RL algorithms: deep Q-networks (DQN) \cite{mnih2013playing}, Advantage-Actor-Critic (A2C) \cite{sutton2018reinforcement}, and proximal policy optimization (PP0) \cite{schulman2017proximal}. DQN were first used to achieve superhuman performance an Atari games \cite{mnih2013playing}, one of the first big successes of deep-RL. A2C and PPO came later, and are popular choices in contemporary research, having strong performance and relatively easy implementations. Figure~\ref{taus-RL} shows the training curves of these algorithms when used on a taxi moving in $\mathcal{M}$ along with the baseline optimal $\tau$. The results are encouraging; the RL-taxis learn policies that give close to optimal $\tau$ on all cities. A2C performs best on all cities, but takes the longest to train. PPO on the other hand trains quickest, while the DQN is slow and unstable. The hyperparameters used for the algorithms are described in Appendix~\ref{appendix:ho}. 

%%%%%%%%%%%%%%%%%%%%%%%%%%%%%%%%%%%%%%%%%%%%%%%%%%%%%%%%%%%%%%%%%%%%%

\subsection{N taxis}
\begin{figure*}
  \includegraphics[width= \linewidth]{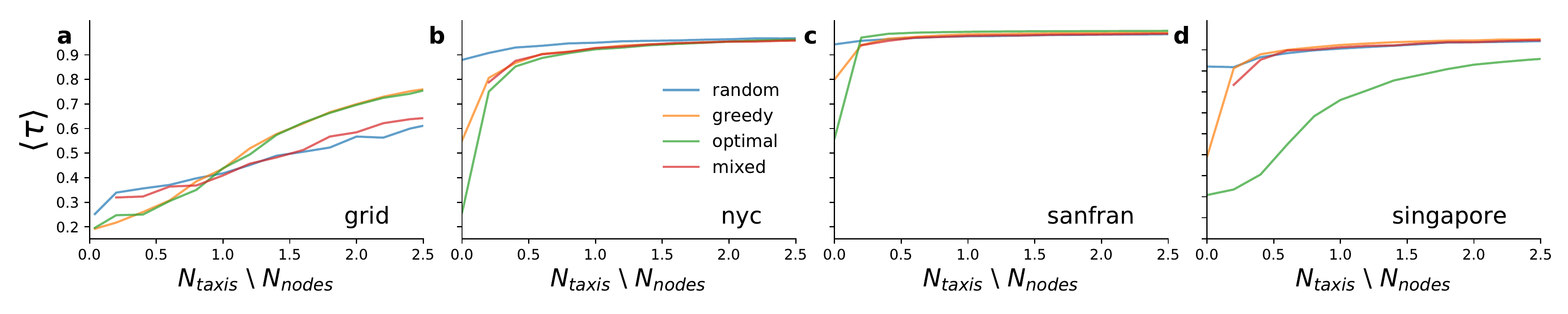}
  \caption{Mean idle time of a population of taxis following deterministic policies discussed in the main text. Note the mean is with respect to the population; just one simulation of length $T = 8640$ (this $T$ is the same for all simulations run in the paper) was run for each data point. The population size $N_{taxis}$ is plotted relative to the number of nodes in the graph $N_{nodes}$ (values for $N_{nodes}$ are reported in Appendix~\ref{appendix:ds}), which is a measure of the density of taxis on the graph. For low densities, the optimal (optimal for the single agent case) $\pi^*$ policy has low idle times. But this performance degrades at higher densities.}
  \label{taus-multi-agent}
\end{figure*}

Next we study the $N > 1$ taxi regime. We first explore if $\pi^*$ -- which, recall, the optimal policy when a \textit{single} taxi moves on the street network $S$ -- is useful in this case. We suspect that in the low density regime, when the probability of two taxis being at the same node is small, that $N$ taxis moving according to $\pi^*$ might still lead to low $\tau$. In Figure~\ref{taus-multi-agent} we test this by showing the mean idle time of a population of taxis moving in $\mathcal{M}$ according to $\pi^*$ versus $N_{taxis} /N_{nodes}$ -- a measure of density. We also show $\langle \tau \rangle$ for three other populations: a population of random taxis, a population of greedy taxis, and a mixed population of equal numbers of $\pi^*$-taxis, greedy taxis, and random taxis. Interestingly, for the grid graph $\pi^*$-taxis and greedy taxis perform worse than random taxis when the density is large. In Singapore, the $\pi^*$ population performs better than the other population for all densities, while for the remaining cities all taxi types plateau at large inefficiencies at large densities. These results show there is much room for improvement for an RL-taxi.

For simplicity, we restrict our analysis to the easier case of just one RL-taxi in a population of regular taxis (taxis following a deterministic policy). This lets us avoid the much more difficult multi-agent RL problem, in which game-theoretic complexities arise. For example, because intelligent agents can change their behavior, the idea of a static optimal policy becomes ill-defined (the policy which beats opponent $A$ at some stage during training might later become ineffective as $A$ changes its policy during learning, and so on). More sophisticated algorithms are needed to train agents capable of handling these complications, which we leave for future work.

%%%%%% Figure 3

%%%%%% Figure 4
\begin{figure*}
  \includegraphics[width= \linewidth]{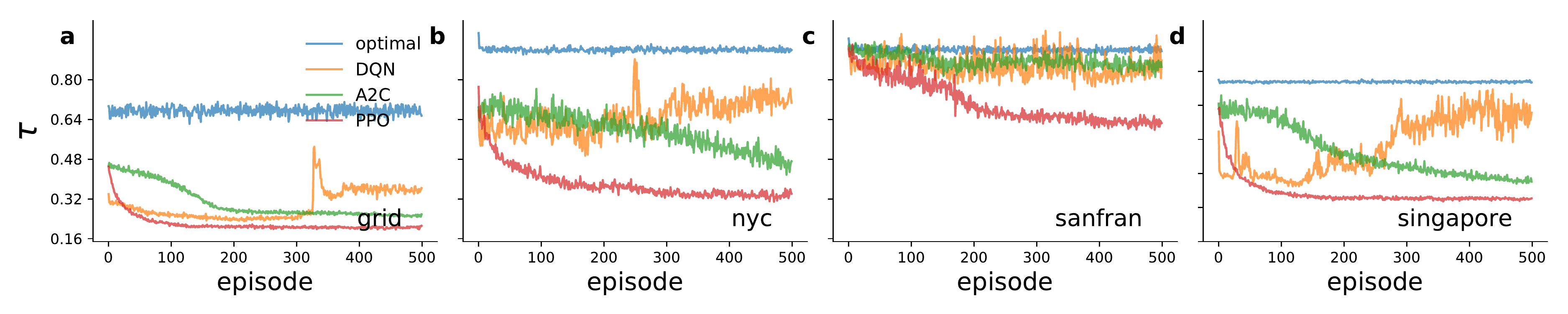}
  \caption{Training curves for all a single RL-taxi operating in a population of $N_{taxis} = 2 N_{nodes}$ taxis following the $\pi^*$ policy. The RL-taxis learns to significantly outperform the other taxis.}
  \label{taus-multi-agent-RL}
\end{figure*}

%Next we test how RL-taxis fare in a population of regular taxis. For simplicity, we consider just one RL-taxi in a population of regular taxis (taxis following a deterministic policy). This lets us avoid the much more difficult multi-agent RL problem, in which game-theoretic complexities arise; because intelligent agents can change their behavior, the idea of a static optimal policy becomes ill-defined (the policy which beats opponent $A$ at some stage during training might later become ineffective as $A$ changes its policy during learning, and so on). More sophisticated algorithms are needed to train agents capable of handling these complications, which we leave for future work.

\begin{figure*}
  \includegraphics[width= \linewidth]{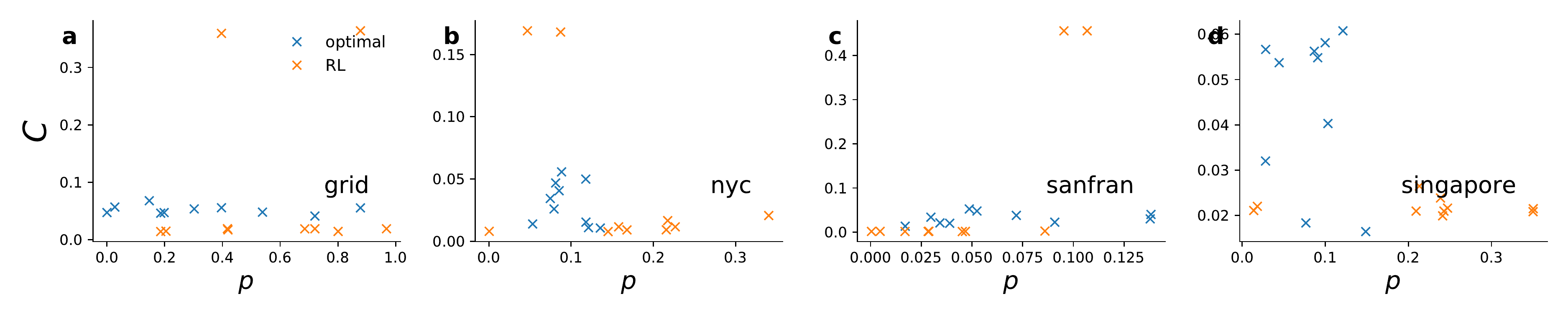}
  \caption{Fraction of time taxis spend at each node $C_i$ versus that node's passenger generation rate $p_i$ for all cities. The RL-taxi was trained using the PPO algorithm with hyper parameters given in Appendix~\ref{appendix:ho}. Only the $10$ largest $C_i$ are shown.}
  \label{optimal-policies}
\end{figure*}

Figure~\ref{taus-multi-agent-RL} shows how a single RL-taxi out performs in a population of $\pi^*$-taxis for all the cities when the density is $N_{taxis} / N_{nodes}$ = 2. Similar results are found when the RL-taxi operates in a population of greedy taxis and random taxis, and for other values of the density. We were curious what strategy the $RL$-taxi followed to beat the other taxis. To this end, in Figure~\ref{optimal-policies} we plotted the relative amount of times the $RL$-taxi (trained using PPO) spends at each node, $C^{RL}_i$, as well as that of the optimal cab $C^{optimal}_i$ versus the passenger generation rate $p_i$ for each city (ideally we would plot the $C_i$ over the street networks too, but it was challenging to plot both the networks along with $C_i$ and the passenger generation rates $p_i$). For all but the Singapore street network, we see that the RL-taxi learns to exploit a few nodes that are not very popular in terms of $p_i$, but are visited less than the other cabs. For Singapore however, we see the $RL$-taxi concentrates on a few popular nodes. 

%%%%%% Figure 5

%%%%%%%%%%%%%%%%%%%%%%%%%%%%%%%%%%%%%%%%%%%%%%%%%%%%%%%%%%%%%%%%%%%%%%%%%%

\section{Discussion}
Our main goal was to explore if RL could be useful for minimizing taxi idle times. To achieve this, we studied taxis moving in two toy environments: a single RL-taxi moving on real-world street networks with realistic passengers demand, and a single RL-taxi moving in a population of regular taxis moving with simple rules. In these idealized settings, we found that indeed RL-taxis learned to minimize their idle times; in the first setting, the PPO algorithms achieved close to optimal behavior in each city studied, and in the second it achieved significantly lower $\tau$ (as compared with the other non-RL taxis). Previous work \cite{han2016routing} studied the first case, but with basic $Q$-learning (non-deep) RL and on only one city. And to our knowledge, no other work has studied our multi-agent set-up. Taken together, our findings indicate that RL could be useful in improving the efficiency of real-world taxis. They are also potentially useful to other aspects of taxi research, such as route planning when serving passengers \cite{li2009hierarchical,koeners2011improving} (as opposed to route planning when searching for passengers) or mobile sensing \cite{o2019quantifying,o2019urban,skordylis2011efficient,piran2011total}. 

The most pertinent direction for future work would be relax the idealizations we made in our study. For example, a more realistic model for the non-RL taxis, such as the Intelligent Driver Model \cite{treiber2000congested} could be used. The same is the true for the passenger demand, which we characterized with a simple Poisson process; real passenger demand most likely has temporal variations -- a complication which could potentially be tackled with distributional RL \cite{bellemare2017distributional}. Betters models for pricing (which we ignored here), and for inter-taxi competition (which we modeled as a simple coin toss) would also be interesting. Finally, a fully multi-agent RL study could be executed, in which a fleet of RL-taxis are trained simultaneously. Effective algorithms for multi-agent RL have recently been developed, which could prove fruitful for this purpose \cite{lowe2017multi,rashid2018qmix,sunehag2017value}.

%%%%%%%%%%%%%%%%%%%%%%%%%%%%%%%%%%%%%%%%%%%%%%%%%%%%%%%%%%%%%%%%%%%%%%%%%%

\bibliographystyle{apsrev}

\begin{thebibliography}{31}
\expandafter\ifx\csname natexlab\endcsname\relax\def\natexlab#1{#1}\fi
\expandafter\ifx\csname bibnamefont\endcsname\relax
  \def\bibnamefont#1{#1}\fi
\expandafter\ifx\csname bibfnamefont\endcsname\relax
  \def\bibfnamefont#1{#1}\fi
\expandafter\ifx\csname citenamefont\endcsname\relax
  \def\citenamefont#1{#1}\fi
\expandafter\ifx\csname url\endcsname\relax
  \def\url#1{\texttt{#1}}\fi
\expandafter\ifx\csname urlprefix\endcsname\relax\def\urlprefix{URL }\fi
\providecommand{\bibinfo}[2]{#2}
\providecommand{\eprint}[2][]{\url{#2}}

\bibitem[{\citenamefont{Schaller}(2017)}]{schaller2017}
\bibinfo{author}{\bibfnamefont{B.}~\bibnamefont{Schaller}},
  \bibinfo{journal}{Schaller Consulting, December} \textbf{\bibinfo{volume}{1}}
  (\bibinfo{year}{2017}).

\bibitem[{\citenamefont{Powell et~al.}(2011)\citenamefont{Powell, Huang,
  Bastani, and Ji}}]{powell2011towards}
\bibinfo{author}{\bibfnamefont{J.~W.} \bibnamefont{Powell}},
  \bibinfo{author}{\bibfnamefont{Y.}~\bibnamefont{Huang}},
  \bibinfo{author}{\bibfnamefont{F.}~\bibnamefont{Bastani}}, \bibnamefont{and}
  \bibinfo{author}{\bibfnamefont{M.}~\bibnamefont{Ji}}, in
  \emph{\bibinfo{booktitle}{International Symposium on Spatial and Temporal
  Databases}} (\bibinfo{organization}{Springer}, \bibinfo{year}{2011}), pp.
  \bibinfo{pages}{242--260}.

\bibitem[{\citenamefont{Zhang et~al.}(2014)\citenamefont{Zhang, He, Lin, Munir,
  and Stankovic}}]{zhang2014online}
\bibinfo{author}{\bibfnamefont{D.}~\bibnamefont{Zhang}},
  \bibinfo{author}{\bibfnamefont{T.}~\bibnamefont{He}},
  \bibinfo{author}{\bibfnamefont{S.}~\bibnamefont{Lin}},
  \bibinfo{author}{\bibfnamefont{S.}~\bibnamefont{Munir}}, \bibnamefont{and}
  \bibinfo{author}{\bibfnamefont{J.~A.} \bibnamefont{Stankovic}},
  \bibinfo{journal}{IEEE Transactions on Parallel and Distributed Systems}
  \textbf{\bibinfo{volume}{26}}, \bibinfo{pages}{3122} (\bibinfo{year}{2014}).

\bibitem[{\citenamefont{Qu et~al.}(2014)\citenamefont{Qu, Zhu, Liu, Liu, and
  Xiong}}]{qu2014cost}
\bibinfo{author}{\bibfnamefont{M.}~\bibnamefont{Qu}},
  \bibinfo{author}{\bibfnamefont{H.}~\bibnamefont{Zhu}},
  \bibinfo{author}{\bibfnamefont{J.}~\bibnamefont{Liu}},
  \bibinfo{author}{\bibfnamefont{G.}~\bibnamefont{Liu}}, \bibnamefont{and}
  \bibinfo{author}{\bibfnamefont{H.}~\bibnamefont{Xiong}}, in
  \emph{\bibinfo{booktitle}{Proceedings of the 20th ACM SIGKDD international
  conference on Knowledge discovery and data mining}}
  (\bibinfo{organization}{ACM}, \bibinfo{year}{2014}), pp.
  \bibinfo{pages}{45--54}.

\bibitem[{\citenamefont{Dong et~al.}(2014)\citenamefont{Dong, Zhang, Dong,
  Chen, and Rao}}]{dong2014recommend}
\bibinfo{author}{\bibfnamefont{H.}~\bibnamefont{Dong}},
  \bibinfo{author}{\bibfnamefont{X.}~\bibnamefont{Zhang}},
  \bibinfo{author}{\bibfnamefont{Y.}~\bibnamefont{Dong}},
  \bibinfo{author}{\bibfnamefont{C.}~\bibnamefont{Chen}}, \bibnamefont{and}
  \bibinfo{author}{\bibfnamefont{F.}~\bibnamefont{Rao}}, in
  \emph{\bibinfo{booktitle}{17th International IEEE Conference on Intelligent
  Transportation Systems (ITSC)}} (\bibinfo{organization}{IEEE},
  \bibinfo{year}{2014}), pp. \bibinfo{pages}{2003--2008}.

\bibitem[{\citenamefont{Huang et~al.}(2014)\citenamefont{Huang, Huangfu, Sun,
  Li, Zhao, Cheng, and Song}}]{huang2014backward}
\bibinfo{author}{\bibfnamefont{J.}~\bibnamefont{Huang}},
  \bibinfo{author}{\bibfnamefont{X.}~\bibnamefont{Huangfu}},
  \bibinfo{author}{\bibfnamefont{H.}~\bibnamefont{Sun}},
  \bibinfo{author}{\bibfnamefont{H.}~\bibnamefont{Li}},
  \bibinfo{author}{\bibfnamefont{P.}~\bibnamefont{Zhao}},
  \bibinfo{author}{\bibfnamefont{H.}~\bibnamefont{Cheng}}, \bibnamefont{and}
  \bibinfo{author}{\bibfnamefont{Q.}~\bibnamefont{Song}},
  \bibinfo{journal}{IEEE transactions on Knowledge and Data Engineering}
  \textbf{\bibinfo{volume}{27}}, \bibinfo{pages}{46} (\bibinfo{year}{2014}).

\bibitem[{\citenamefont{Yu et~al.}(2019)\citenamefont{Yu, Gao, Hu, and
  Park}}]{yu2019markov}
\bibinfo{author}{\bibfnamefont{X.}~\bibnamefont{Yu}},
  \bibinfo{author}{\bibfnamefont{S.}~\bibnamefont{Gao}},
  \bibinfo{author}{\bibfnamefont{X.}~\bibnamefont{Hu}}, \bibnamefont{and}
  \bibinfo{author}{\bibfnamefont{H.}~\bibnamefont{Park}},
  \bibinfo{journal}{Transportation Research Part B: Methodological}
  \textbf{\bibinfo{volume}{121}}, \bibinfo{pages}{114} (\bibinfo{year}{2019}).

\bibitem[{\citenamefont{Ng et~al.}(2006)\citenamefont{Ng, Coates, Diel,
  Ganapathi, Schulte, Tse, Berger, and Liang}}]{ng2006autonomous}
\bibinfo{author}{\bibfnamefont{A.~Y.} \bibnamefont{Ng}},
  \bibinfo{author}{\bibfnamefont{A.}~\bibnamefont{Coates}},
  \bibinfo{author}{\bibfnamefont{M.}~\bibnamefont{Diel}},
  \bibinfo{author}{\bibfnamefont{V.}~\bibnamefont{Ganapathi}},
  \bibinfo{author}{\bibfnamefont{J.}~\bibnamefont{Schulte}},
  \bibinfo{author}{\bibfnamefont{B.}~\bibnamefont{Tse}},
  \bibinfo{author}{\bibfnamefont{E.}~\bibnamefont{Berger}}, \bibnamefont{and}
  \bibinfo{author}{\bibfnamefont{E.}~\bibnamefont{Liang}}, in
  \emph{\bibinfo{booktitle}{Experimental robotics IX}}
  (\bibinfo{publisher}{Springer}, \bibinfo{year}{2006}), pp.
  \bibinfo{pages}{363--372}.

\bibitem[{\citenamefont{Abbeel et~al.}(2007)\citenamefont{Abbeel, Coates,
  Quigley, and Ng}}]{abbeel2007application}
\bibinfo{author}{\bibfnamefont{P.}~\bibnamefont{Abbeel}},
  \bibinfo{author}{\bibfnamefont{A.}~\bibnamefont{Coates}},
  \bibinfo{author}{\bibfnamefont{M.}~\bibnamefont{Quigley}}, \bibnamefont{and}
  \bibinfo{author}{\bibfnamefont{A.~Y.} \bibnamefont{Ng}}, in
  \emph{\bibinfo{booktitle}{Advances in neural information processing systems}}
  (\bibinfo{year}{2007}), pp. \bibinfo{pages}{1--8}.

\bibitem[{\citenamefont{Silver et~al.}(2017)\citenamefont{Silver,
  Schrittwieser, Simonyan, Antonoglou, Huang, Guez, Hubert, Baker, Lai, Bolton
  et~al.}}]{silver2017mastering}
\bibinfo{author}{\bibfnamefont{D.}~\bibnamefont{Silver}},
  \bibinfo{author}{\bibfnamefont{J.}~\bibnamefont{Schrittwieser}},
  \bibinfo{author}{\bibfnamefont{K.}~\bibnamefont{Simonyan}},
  \bibinfo{author}{\bibfnamefont{I.}~\bibnamefont{Antonoglou}},
  \bibinfo{author}{\bibfnamefont{A.}~\bibnamefont{Huang}},
  \bibinfo{author}{\bibfnamefont{A.}~\bibnamefont{Guez}},
  \bibinfo{author}{\bibfnamefont{T.}~\bibnamefont{Hubert}},
  \bibinfo{author}{\bibfnamefont{L.}~\bibnamefont{Baker}},
  \bibinfo{author}{\bibfnamefont{M.}~\bibnamefont{Lai}},
  \bibinfo{author}{\bibfnamefont{A.}~\bibnamefont{Bolton}},
  \bibnamefont{et~al.}, \bibinfo{journal}{Nature}
  \textbf{\bibinfo{volume}{550}}, \bibinfo{pages}{354} (\bibinfo{year}{2017}).

\bibitem[{\citenamefont{Andrychowicz et~al.}(2018)\citenamefont{Andrychowicz,
  Baker, Chociej, Jozefowicz, McGrew, Pachocki, Petron, Plappert, Powell, Ray
  et~al.}}]{andrychowicz2018learning}
\bibinfo{author}{\bibfnamefont{M.}~\bibnamefont{Andrychowicz}},
  \bibinfo{author}{\bibfnamefont{B.}~\bibnamefont{Baker}},
  \bibinfo{author}{\bibfnamefont{M.}~\bibnamefont{Chociej}},
  \bibinfo{author}{\bibfnamefont{R.}~\bibnamefont{Jozefowicz}},
  \bibinfo{author}{\bibfnamefont{B.}~\bibnamefont{McGrew}},
  \bibinfo{author}{\bibfnamefont{J.}~\bibnamefont{Pachocki}},
  \bibinfo{author}{\bibfnamefont{A.}~\bibnamefont{Petron}},
  \bibinfo{author}{\bibfnamefont{M.}~\bibnamefont{Plappert}},
  \bibinfo{author}{\bibfnamefont{G.}~\bibnamefont{Powell}},
  \bibinfo{author}{\bibfnamefont{A.}~\bibnamefont{Ray}}, \bibnamefont{et~al.},
  \bibinfo{journal}{arXiv preprint arXiv:1808.00177}  (\bibinfo{year}{2018}).

\bibitem[{\citenamefont{Jindal et~al.}(2018)\citenamefont{Jindal, Qin, Chen,
  Nokleby, and Ye}}]{jindal2018optimizing}
\bibinfo{author}{\bibfnamefont{I.}~\bibnamefont{Jindal}},
  \bibinfo{author}{\bibfnamefont{Z.~T.} \bibnamefont{Qin}},
  \bibinfo{author}{\bibfnamefont{X.}~\bibnamefont{Chen}},
  \bibinfo{author}{\bibfnamefont{M.}~\bibnamefont{Nokleby}}, \bibnamefont{and}
  \bibinfo{author}{\bibfnamefont{J.}~\bibnamefont{Ye}}, in
  \emph{\bibinfo{booktitle}{2018 IEEE International Conference on Big Data (Big
  Data)}} (\bibinfo{organization}{IEEE}, \bibinfo{year}{2018}), pp.
  \bibinfo{pages}{1417--1426}.

\bibitem[{\citenamefont{Oda and Joe-Wong}(2018)}]{oda2018movi}
\bibinfo{author}{\bibfnamefont{T.}~\bibnamefont{Oda}} \bibnamefont{and}
  \bibinfo{author}{\bibfnamefont{C.}~\bibnamefont{Joe-Wong}}, in
  \emph{\bibinfo{booktitle}{IEEE INFOCOM 2018-IEEE Conference on Computer
  Communications}} (\bibinfo{organization}{IEEE}, \bibinfo{year}{2018}), pp.
  \bibinfo{pages}{2708--2716}.

\bibitem[{\citenamefont{Shi et~al.}(2018)\citenamefont{Shi, Ding, Errapotu,
  Yue, Xu, Zhou, and Pan}}]{shi2018deep}
\bibinfo{author}{\bibfnamefont{D.}~\bibnamefont{Shi}},
  \bibinfo{author}{\bibfnamefont{J.}~\bibnamefont{Ding}},
  \bibinfo{author}{\bibfnamefont{S.~M.} \bibnamefont{Errapotu}},
  \bibinfo{author}{\bibfnamefont{H.}~\bibnamefont{Yue}},
  \bibinfo{author}{\bibfnamefont{W.}~\bibnamefont{Xu}},
  \bibinfo{author}{\bibfnamefont{X.}~\bibnamefont{Zhou}}, \bibnamefont{and}
  \bibinfo{author}{\bibfnamefont{M.}~\bibnamefont{Pan}}, in
  \emph{\bibinfo{booktitle}{2018 IEEE Global Communications Conference
  (GLOBECOM)}} (\bibinfo{organization}{IEEE}, \bibinfo{year}{2018}), pp.
  \bibinfo{pages}{1--7}.

\bibitem[{\citenamefont{Han et~al.}(2016)\citenamefont{Han, Senellart, Bressan,
  and Wu}}]{han2016routing}
\bibinfo{author}{\bibfnamefont{M.}~\bibnamefont{Han}},
  \bibinfo{author}{\bibfnamefont{P.}~\bibnamefont{Senellart}},
  \bibinfo{author}{\bibfnamefont{S.}~\bibnamefont{Bressan}}, \bibnamefont{and}
  \bibinfo{author}{\bibfnamefont{H.}~\bibnamefont{Wu}}, in
  \emph{\bibinfo{booktitle}{Proceedings of the 25th ACM International on
  Conference on Information and Knowledge Management}}
  (\bibinfo{organization}{ACM}, \bibinfo{year}{2016}), pp.
  \bibinfo{pages}{2421--2424}.

\bibitem[{\citenamefont{Sutton and Barto}(2018)}]{sutton2018reinforcement}
\bibinfo{author}{\bibfnamefont{R.~S.} \bibnamefont{Sutton}} \bibnamefont{and}
  \bibinfo{author}{\bibfnamefont{A.~G.} \bibnamefont{Barto}},
  \emph{\bibinfo{title}{Reinforcement learning: An introduction}}
  (\bibinfo{publisher}{MIT press}, \bibinfo{year}{2018}).

\bibitem[{\citenamefont{Mnih et~al.}(2013)\citenamefont{Mnih, Kavukcuoglu,
  Silver, Graves, Antonoglou, Wierstra, and Riedmiller}}]{mnih2013playing}
\bibinfo{author}{\bibfnamefont{V.}~\bibnamefont{Mnih}},
  \bibinfo{author}{\bibfnamefont{K.}~\bibnamefont{Kavukcuoglu}},
  \bibinfo{author}{\bibfnamefont{D.}~\bibnamefont{Silver}},
  \bibinfo{author}{\bibfnamefont{A.}~\bibnamefont{Graves}},
  \bibinfo{author}{\bibfnamefont{I.}~\bibnamefont{Antonoglou}},
  \bibinfo{author}{\bibfnamefont{D.}~\bibnamefont{Wierstra}}, \bibnamefont{and}
  \bibinfo{author}{\bibfnamefont{M.}~\bibnamefont{Riedmiller}},
  \bibinfo{journal}{arXiv preprint arXiv:1312.5602}  (\bibinfo{year}{2013}).

\bibitem[{\citenamefont{Schulman et~al.}(2017)\citenamefont{Schulman, Wolski,
  Dhariwal, Radford, and Klimov}}]{schulman2017proximal}
\bibinfo{author}{\bibfnamefont{J.}~\bibnamefont{Schulman}},
  \bibinfo{author}{\bibfnamefont{F.}~\bibnamefont{Wolski}},
  \bibinfo{author}{\bibfnamefont{P.}~\bibnamefont{Dhariwal}},
  \bibinfo{author}{\bibfnamefont{A.}~\bibnamefont{Radford}}, \bibnamefont{and}
  \bibinfo{author}{\bibfnamefont{O.}~\bibnamefont{Klimov}},
  \bibinfo{journal}{arXiv preprint arXiv:1707.06347}  (\bibinfo{year}{2017}).

\bibitem[{\citenamefont{Li et~al.}(2009)\citenamefont{Li, Zeng, Yang, and
  Zhang}}]{li2009hierarchical}
\bibinfo{author}{\bibfnamefont{Q.}~\bibnamefont{Li}},
  \bibinfo{author}{\bibfnamefont{Z.}~\bibnamefont{Zeng}},
  \bibinfo{author}{\bibfnamefont{B.}~\bibnamefont{Yang}}, \bibnamefont{and}
  \bibinfo{author}{\bibfnamefont{T.}~\bibnamefont{Zhang}}, in
  \emph{\bibinfo{booktitle}{2009 17th International Conference on
  Geoinformatics}} (\bibinfo{organization}{IEEE}, \bibinfo{year}{2009}), pp.
  \bibinfo{pages}{1--5}.

\bibitem[{\citenamefont{Koeners et~al.}(2011)\citenamefont{Koeners, Stout, and
  Rademaker}}]{koeners2011improving}
\bibinfo{author}{\bibfnamefont{G.}~\bibnamefont{Koeners}},
  \bibinfo{author}{\bibfnamefont{E.}~\bibnamefont{Stout}}, \bibnamefont{and}
  \bibinfo{author}{\bibfnamefont{R.}~\bibnamefont{Rademaker}}, in
  \emph{\bibinfo{booktitle}{2011 IEEE/AIAA 30th Digital Avionics Systems
  Conference}} (\bibinfo{organization}{IEEE}, \bibinfo{year}{2011}), pp.
  \bibinfo{pages}{2C3--1}.

\bibitem[{\citenamefont{O’Keeffe et~al.}(2019)\citenamefont{O’Keeffe,
  Anjomshoaa, Strogatz, Santi, and Ratti}}]{o2019quantifying}
\bibinfo{author}{\bibfnamefont{K.~P.} \bibnamefont{O’Keeffe}},
  \bibinfo{author}{\bibfnamefont{A.}~\bibnamefont{Anjomshoaa}},
  \bibinfo{author}{\bibfnamefont{S.~H.} \bibnamefont{Strogatz}},
  \bibinfo{author}{\bibfnamefont{P.}~\bibnamefont{Santi}}, \bibnamefont{and}
  \bibinfo{author}{\bibfnamefont{C.}~\bibnamefont{Ratti}},
  \bibinfo{journal}{Proceedings of the National Academy of Sciences}
  \textbf{\bibinfo{volume}{116}}, \bibinfo{pages}{12752}
  (\bibinfo{year}{2019}).

\bibitem[{\citenamefont{O'Keeffe et~al.}(2019)\citenamefont{O'Keeffe, Santi,
  Wang, and Ratti}}]{o2019urban}
\bibinfo{author}{\bibfnamefont{K.}~\bibnamefont{O'Keeffe}},
  \bibinfo{author}{\bibfnamefont{P.}~\bibnamefont{Santi}},
  \bibinfo{author}{\bibfnamefont{B.}~\bibnamefont{Wang}}, \bibnamefont{and}
  \bibinfo{author}{\bibfnamefont{C.}~\bibnamefont{Ratti}},
  \bibinfo{journal}{arXiv preprint arXiv:1901.08678}  (\bibinfo{year}{2019}).

\bibitem[{\citenamefont{Skordylis and Trigoni}(2011)}]{skordylis2011efficient}
\bibinfo{author}{\bibfnamefont{A.}~\bibnamefont{Skordylis}} \bibnamefont{and}
  \bibinfo{author}{\bibfnamefont{N.}~\bibnamefont{Trigoni}},
  \bibinfo{journal}{IEEE Transactions on Intelligent Transportation Systems}
  \textbf{\bibinfo{volume}{12}}, \bibinfo{pages}{680} (\bibinfo{year}{2011}).

\bibitem[{\citenamefont{Piran et~al.}(2011)\citenamefont{Piran, Murthy, Babu,
  and Ahvar}}]{piran2011total}
\bibinfo{author}{\bibfnamefont{M.~J.} \bibnamefont{Piran}},
  \bibinfo{author}{\bibfnamefont{G.~R.} \bibnamefont{Murthy}},
  \bibinfo{author}{\bibfnamefont{G.~P.} \bibnamefont{Babu}}, \bibnamefont{and}
  \bibinfo{author}{\bibfnamefont{E.}~\bibnamefont{Ahvar}}, in
  \emph{\bibinfo{booktitle}{2011 Third International Conference on
  Computational Intelligence, Modelling \& Simulation}}
  (\bibinfo{organization}{IEEE}, \bibinfo{year}{2011}), pp.
  \bibinfo{pages}{388--393}.

\bibitem[{\citenamefont{Treiber et~al.}(2000)\citenamefont{Treiber, Hennecke,
  and Helbing}}]{treiber2000congested}
\bibinfo{author}{\bibfnamefont{M.}~\bibnamefont{Treiber}},
  \bibinfo{author}{\bibfnamefont{A.}~\bibnamefont{Hennecke}}, \bibnamefont{and}
  \bibinfo{author}{\bibfnamefont{D.}~\bibnamefont{Helbing}},
  \bibinfo{journal}{Physical review E} \textbf{\bibinfo{volume}{62}},
  \bibinfo{pages}{1805} (\bibinfo{year}{2000}).

\bibitem[{\citenamefont{Bellemare et~al.}(2017)\citenamefont{Bellemare, Dabney,
  and Munos}}]{bellemare2017distributional}
\bibinfo{author}{\bibfnamefont{M.~G.} \bibnamefont{Bellemare}},
  \bibinfo{author}{\bibfnamefont{W.}~\bibnamefont{Dabney}}, \bibnamefont{and}
  \bibinfo{author}{\bibfnamefont{R.}~\bibnamefont{Munos}}, in
  \emph{\bibinfo{booktitle}{Proceedings of the 34th International Conference on
  Machine Learning-Volume 70}} (\bibinfo{organization}{JMLR. org},
  \bibinfo{year}{2017}), pp. \bibinfo{pages}{449--458}.

\bibitem[{\citenamefont{Lowe et~al.}(2017)\citenamefont{Lowe, Wu, Tamar, Harb,
  Abbeel, and Mordatch}}]{lowe2017multi}
\bibinfo{author}{\bibfnamefont{R.}~\bibnamefont{Lowe}},
  \bibinfo{author}{\bibfnamefont{Y.}~\bibnamefont{Wu}},
  \bibinfo{author}{\bibfnamefont{A.}~\bibnamefont{Tamar}},
  \bibinfo{author}{\bibfnamefont{J.}~\bibnamefont{Harb}},
  \bibinfo{author}{\bibfnamefont{O.~P.} \bibnamefont{Abbeel}},
  \bibnamefont{and} \bibinfo{author}{\bibfnamefont{I.}~\bibnamefont{Mordatch}},
  in \emph{\bibinfo{booktitle}{Advances in Neural Information Processing
  Systems}} (\bibinfo{year}{2017}), pp. \bibinfo{pages}{6379--6390}.

\bibitem[{\citenamefont{Rashid et~al.}(2018)\citenamefont{Rashid, Samvelyan,
  De~Witt, Farquhar, Foerster, and Whiteson}}]{rashid2018qmix}
\bibinfo{author}{\bibfnamefont{T.}~\bibnamefont{Rashid}},
  \bibinfo{author}{\bibfnamefont{M.}~\bibnamefont{Samvelyan}},
  \bibinfo{author}{\bibfnamefont{C.~S.} \bibnamefont{De~Witt}},
  \bibinfo{author}{\bibfnamefont{G.}~\bibnamefont{Farquhar}},
  \bibinfo{author}{\bibfnamefont{J.}~\bibnamefont{Foerster}}, \bibnamefont{and}
  \bibinfo{author}{\bibfnamefont{S.}~\bibnamefont{Whiteson}},
  \bibinfo{journal}{arXiv preprint arXiv:1803.11485}  (\bibinfo{year}{2018}).

\bibitem[{\citenamefont{Sunehag et~al.}(2017)\citenamefont{Sunehag, Lever,
  Gruslys, Czarnecki, Zambaldi, Jaderberg, Lanctot, Sonnerat, Leibo, Tuyls
  et~al.}}]{sunehag2017value}
\bibinfo{author}{\bibfnamefont{P.}~\bibnamefont{Sunehag}},
  \bibinfo{author}{\bibfnamefont{G.}~\bibnamefont{Lever}},
  \bibinfo{author}{\bibfnamefont{A.}~\bibnamefont{Gruslys}},
  \bibinfo{author}{\bibfnamefont{W.~M.} \bibnamefont{Czarnecki}},
  \bibinfo{author}{\bibfnamefont{V.}~\bibnamefont{Zambaldi}},
  \bibinfo{author}{\bibfnamefont{M.}~\bibnamefont{Jaderberg}},
  \bibinfo{author}{\bibfnamefont{M.}~\bibnamefont{Lanctot}},
  \bibinfo{author}{\bibfnamefont{N.}~\bibnamefont{Sonnerat}},
  \bibinfo{author}{\bibfnamefont{J.~Z.} \bibnamefont{Leibo}},
  \bibinfo{author}{\bibfnamefont{K.}~\bibnamefont{Tuyls}},
  \bibnamefont{et~al.}, \bibinfo{journal}{arXiv preprint arXiv:1706.05296}
  (\bibinfo{year}{2017}).

\bibitem[{san()}]{sanfran_data_set}
\emph{\bibinfo{title}{san fran dataset}},
  \bibinfo{howpublished}{\url{http://crawdad.org/dartmouth/campus/20090909}},
  \bibinfo{note}{accessed: 01/04/2016}.

\bibitem[{\citenamefont{Santi et~al.}(2014)\citenamefont{Santi, Resta, Szell,
  Sobolevsky, Strogatz, and Ratti}}]{santi2014}
\bibinfo{author}{\bibfnamefont{P.}~\bibnamefont{Santi}},
  \bibinfo{author}{\bibfnamefont{G.}~\bibnamefont{Resta}},
  \bibinfo{author}{\bibfnamefont{M.}~\bibnamefont{Szell}},
  \bibinfo{author}{\bibfnamefont{S.}~\bibnamefont{Sobolevsky}},
  \bibinfo{author}{\bibfnamefont{S.~H.} \bibnamefont{Strogatz}},
  \bibnamefont{and} \bibinfo{author}{\bibfnamefont{C.}~\bibnamefont{Ratti}},
  \bibinfo{journal}{Proceedings of the National Academy of Sciences}
  \textbf{\bibinfo{volume}{111}}, \bibinfo{pages}{13290}
  (\bibinfo{year}{2014}).

\end{thebibliography}

%\documentclass{article} % For LaTeX2e
%\usepackage{iclr2019_conference,times}

% Optional math commands from https://github.com/goodfeli/dlbook_notation.
%\input{math_commands.tex}

%\usepackage{hyperref}
%\usepackage{url}
%\usepackage{graphicx}
%\usepackage{caption}
%\usepackage{cleveref}
%\usepackage{textgreek}
%\usepackage{xspace}
%\usepackage[binary-units]{siunitx}
%\usepackage{amsmath}
%\usepackage{booktabs}
%\usepackage{listings}
%\usepackage{xcolor}
%\usepackage{verbatim}

%\title{On the Use of ArXiv as a Dataset: Supplemental Information}
%\author{}
%\iclrfinalcopy

%\begin{document}

%\maketitle

\appendix

%%%%%%%%%%%%%%%%%%%%%%%%%%%%%%%%%%%%%%%%%%%%%%%%%%%%%%%%%%%%%%%%%%%%%%%%%%%%%%%%%%%%%%%%%%%%%%%%%%%%%%%%%%%%%%%%%%%%%%%%%%%%%%%%%%%%%%%%%%%%%%

\section{Datasets}
\label{appendix:ds}

\section*{Data sets}
\textbf{Description of datasets}
We used 3 real-world data sets from: New York City (confined to the burough of Manhattan), San Francisco, and Singapore. The New York data set has been obtained from the New York Taxi and Limousine Commission for the year 2011 via a Freedom of Information Act request. The Singapore data set were provided to the MIT SENSEable City Lab by AIT and the Singapore government, respectively. The San Francisco data sets was publicly available \cite{sanfran_data_set}. We note these are the same datasets used in previous studies \cite{santi2014,o2019quantifying} (and so the descriptions of these datasets overlap with those given in those works).

Each data set consists of a set of taxis trips. Each trip $i$ is represented by a GPS coordinate of pickup location $O_i$ and dropoff location $D_i$ (as well as the pickup times and dropoff times). We snap these GPS coordinates to the nearest street segments using OpenStreetMap. We do not however have details on the trajectory of each taxi -- that is, on the intermediary path taken by the taxi when bringing the passenger from $O_i$ to $D_i$. So we need to approximate trajectories. We used two methods for this, one sophisticated, one simple. The sophisticated method was for the Manhattan data set. Here, as was done in \cite{santi2014}, we generated 24 travel time matrices, one for each hour of the day. An element of the matrix $(i,j)$ contains the travel time from intersection $i$ to intersection $j$. Given these matrices, for a particular starting time of the trip, you pick the right matrix for travel time estimation, and compute the shortest time route between origin and destination; that gives an estimation of the trajectory taken for the trip. For the remaining cities, we used the simple method of finding the weighted shortest path between $O_i$ and $D_i$ (where segments were weighted by their length).

The temporal range of the data sets was not uniform. NYC was the most comprehensive, consisting of all the taxi trips in Manhattan starting on 2011.The remaining data sets were for one week. One day's worth of data was also used in the paper: 01/18/11, 05/21/08, 02/21/11, for NYC, San Francisco, Singapore. The sizes of the street networks were $N_{nodes} = (3564,8014,16289)$ in the same order. Training RL-taxis on the entire street networks was prohibitively costly. So we sub-sampled the street networks by selecting a point at the city center and taking all nodes within a distance $R$ of this point. The python commands to find the networks is given in the code block below. The result sub-networks were also smaller that $N_{nodes} = 1000$.

\begin{lstlisting}
import osmnx as ox

def load_sub_graph(city):

    if city == 'nyc':
        point, R = (40.7896239, -73.9598939), 2*10**3
        G = ox.graph_from_point(point, distance=R, network_type='drive')
        
    elif city == 'sanfran':
        R = 1.5*10**3
        G = ox.graph_from_point((37.756986, -122.439705), distance=R, network_type='drive')
        
    elif city == 'singapore':
        R = 6.0*10**3
        G = ox.graph_from_point((1.223439, 103.819684), distance=R,network_type='drive')
        
    return G
\end{lstlisting}
%\caption{Code used to generated the sub street networks}

\begin{figure*}[t!]
  \includegraphics[width= \linewidth]{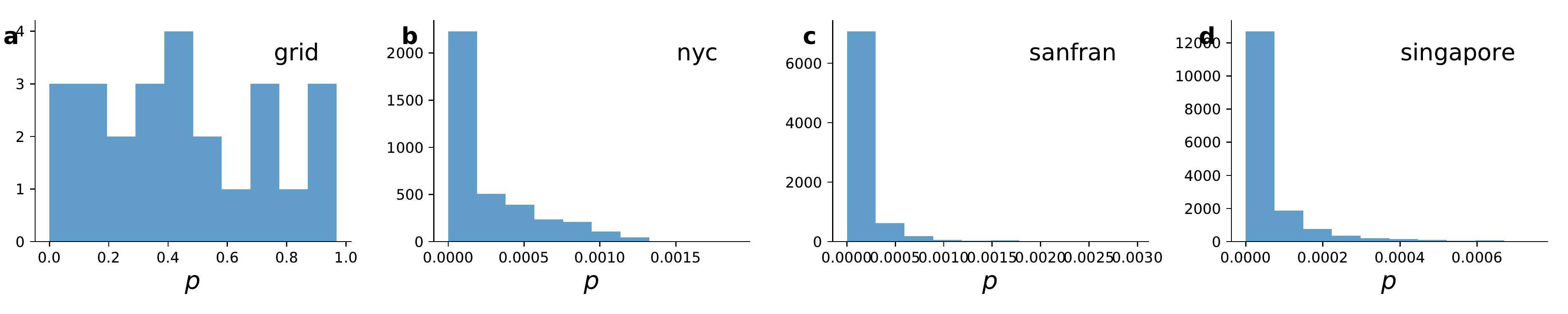}
  \caption{Trip generation probabilities $p_i$ from each city. The grid-world $p_i$ are chosen uniformly at random. The  }
  \label{ps}
\end{figure*}

\textbf{Estimating trip probabilities $p_i$}. We assume passenger trips were generated via a poisson process
\begin{equation}
   \mathbb{P}(N_i = n, t) = \frac{(\lambda_i t)^n}{n!} e^{- \lambda_i t} 
\end{equation}
The probability of at the probability of there being at least one trip generated at node $i$ per time unit is then
\begin{align}
    p_i &= \mathbb{P}(N_i > 0 , t=1) = 1 - \mathbb{P}(N_i = 0 , t=1) \\  
    p_i &= 1 - e^{-\lambda_i t}
\end{align}
We estimated the rate $\lambda_i$ for the $i$-th node by counting the number of trips which start at that node, $C_i$, and divide by the number of times units in a day, $T$ (we take a day as our reference period). The average time taken for a taxi to traverse a segment is $10$ s. So for convenience we measured time in deci-seconds so that taxis move at unit speed. This means $T = 8640$. Figure~\ref{ps} shows the estimated $p_i$.

%%%%%% Figure 1

%%%%%%%%%%%%%%%%%%%%%%%%%%%%%%%%%%%%%%%%%%%%%%%%%%%%%%%%%%%%%%%%%%%%%%%%%%%%%%%%%%%%%%%%%%%%%%%%%%%%%%%%%%%%%%%%%%%%%%%%%%%%%%%%%%%%%%%%%%%%%%

\section{Generalizations to environment $\mathcal{M}$}
\label{appendix:general}
When studying model $\mathcal{M}$ is the main text, we made three simplifications: (i) Passengers trip were generated at each time step with probability $p_i$, and then disappear afterwards (i.e. passenger do not wait at nodes) (ii) Segments were all unit length. We here show these are easily relaxed.

\textbf{Passenger waiting times}. This is easily handled. Assume passengers wait for $\delta$ time-steps after generation. If there is no competition among taxi drivers, this is simply equivalent to increasing the probabilities at each node by a factor of $\delta$, $p_i \rightarrow \delta p_i$. Everything else remains the same. 

\textbf{Variable street lengths}. Non-unit street lengths means that trip generation must be modeled as a Poisson random variables,  as opposed to Bernoulii random variables. This is because a passenger can be picked up at any point along the street segments, whereas before, street segments were treated as one dimensional points. Let $\lambda_i$ be the rate of trip generation .We need to know the probability that the exponential variable is greater than the street length $L$. This is given by complementary CDF of the exponential distribution: ($1 - e^{\lambda x}$). This new form of the fixed point equation is this
\begin{equation}
x_i = (1 - e^{-\lambda_i L_i})(\lambda_i^{-1} - l(e^{\lambda_i L_i} - 1)^{-1}) + e^{-\lambda_i L_i}\left( L_i + \min_{j \in N(i)} x_j \right)    
\end{equation}

\iffalse
\textbf{Non-uniform passenger destinations}.  Let $M_{ij}$ be the probability a passenger who hails a ride at street $i$ is going to street $j$. Let $x_i^n$ be the expected total waiting time over $n$ trips starting from street $i$. Assume unit length streets for simplicity. Then

\begin{align}
\label{unitMulti}
x^0 &= 0 \\
x^n &= P(1 + M^Tx^{n-1}) + Q(A \otimes x^n)
\end{align}

With a proportional policy, this is 

\begin{align}
x^0 &= 0 \\
x^n &= P(1 + M^Tx^{n-1}) + Q(1 + \Pi_{x_n}^T x^n)
\end{align}
\fi

%%%%%%%%%%%%%%%%%%%%%%%%%%%%%%%%%%%%%%%%%%%%%%%%%%%%%%%%%%%%%%%%%%%%%%%%%%%%%%%%%%%%%%%%%%%%%%%%%%%%%%%%%%%%%%%%%%%%%%%%%%%%%%%%%%%%%%%%%%%%

\section{Hyperparameters}
\label{appendix:ho}

We here discuss the hyperparameters for each RL algorithm used. %Implementations of the algorithms in python is available at \footnote{\url{https://github.com/Khev/Qcabs}}.

\textbf{DQN}, The neural network used had three Dense layers of dimension 64. The first two layers had ReLU activations. The final layer had a linear activation. A mean square loss was used with an Adam's optimizer with learning rate $0.01$. The reward discount factor was $0.99$. Following \cite{mnih2013playing}, we used both a behavior and target network. The target network was updated `softly' every learning step according to $\theta_{target} = \tau*\theta_{behavior} + (1 - \tau) \theta_{target}$ with $\tau = 0.01$ where $\theta_{i}$ denotes the parameters in neural network $i$. The DQN-agent had memory size $10^4$ and batch size $32$ was used. During training, actions were selected according to an $\epsilon$-greedy policy with $\epsilon = 0.5$. This value was decayed by factor $\delta = 0.999$ every episode, to a minimum value of $0.05$. Learning was performed every timestep. These values of hyperparameters were found to be optimal (in the sense they produced the lower idle times) on the NYC network. Experiments showed these values were approximately optimal on the other cities too, so these values were frozen throughout the work. See \cite{mnih2013playing} for a review of DQN's and the meaning of these hyperparameters.

\textbf{A2C}. The actor network had three Dense layers of dimension $32$. The first two layers had ReLU activations, while the final layer had softmax. The critic network had the same architecture, except the final layer had dimension $1$ and a linear activation (to return the $Q$-value). Both networks used a Adam's optimizer with learning rate $0.01$. The discount factor was $\gamma = 0.99$. Learning was applied at the end of every episode (in contrast to the DQN, which was done at every timestep in each episode). As for the DQN these values of hyperparameters were found to be optimal (in the sense they produced the lower idle times) on the NYC network and used throughout the work. See \cite{sutton2018reinforcement} for a description of the A2C algorithm.

\textbf{PPO}. The actor network had three Dense layers of dimension 20. The first two layers had $tanh$ activations while the last had a softmax. The critic network is the same as the actor network with a linear activation in the final layer of dimension $1$. An Adam's optimizer was used for both networks. In contrast to the DQN and A2C, different hyperparameter values were used in each city (because experiments showed significantly different results at different values). The hyperparameters were: the learning rate $lr$, loss clipping values $lc$, entropy loss term $c_1$, batch size, and number of epochs used during each learning step (PPO uses more than one gradient step / epoch per learning step). These hyperparameters were found by sampling $20$ times from the following hypergrid:

\begin{lstlisting}
lrs = [0.1,0.01,0.001]
gammas = [0.1,0.5,0.9,0.99]
taus = [0.1]
loss_clips = [0.5,0.2,0.1,0.05,0.01]
c1s = [0.1,0.01,0.001,0.0001]
num_epochs = [2,4,8,16]
batchsize = [16,32,64]
\end{lstlisting}

The optimal values for the grid, NYC, San Francisco, and Singapore street networks were:

\begin{itemize}
    \item $ lr, gamma, loss_{clipping}, c1, batch size, num \; epochs =  0.001, 0.9,  0.1, 0.1, 64, 16 $
    \item $ lr, gamma, loss_{clipping}, c1, batch size, num \; epochs  =  0.001, 0.9,  0.1, 0.1, 64, 16 $
    \item $ lr, gamma, loss_{clipping}, c1, batch size, num \; epochs =  0.001, 0.9, 0.1, 0.1, 64, 16 $
    \item $ lr, gamma, loss_{clipping}, c1, batch size, num \; epochs  =  0.001, 0.99, 0.2, 0.01, 16, 2 $
\end{itemize}

%%%%%%%%%%%%%%%%%%%%%%%%%%%%%%%%%%%%%%%%%%%%%%%%%%%%%%%%%%%%%%%%%%%%%%%%%%%%%%%%%%%%%%%%%%%%%%%%%%%%%%%%%%%%%%%%%%%%%%%%%%%%%%%%%%%%%%%%%%%%%%

\end{document}